%% file: main.tex
\pdfoutput=1
\documentclass[11pt]{article}
\usepackage{graphicx}
\graphicspath{{figs/}}
\usepackage[preprint]{acl}

\usepackage{times}
\usepackage{latexsym}

\usepackage[T1]{fontenc}
\usepackage[utf8]{inputenc}

\usepackage{microtype}
\usepackage{balance}
\usepackage{inconsolata}

\usepackage{multirow}
\usepackage{subfigure}
\usepackage{amsmath}
\usepackage{booktabs}
\usepackage{amssymb}
\usepackage{tabularx}
\usepackage{enumitem}
\usepackage{adjustbox}
\usepackage{hyperref}
\usepackage{graphicx}

\def\model{XRec}

\title{XRec: Large Language Models for Explainable Recommendation}
\author{Qiyao Ma, Xubin Ren, \textbf{Chao Huang}$^{*}$ \\
University of Hong Kong \\
\{martin.qyma, xubinrencs, chaohuang75\}@gmail.com
}

\begin{document}
\maketitle
\begin{abstract}
Recommender systems help users navigate information overload by providing personalized recommendations aligned with their preferences. Collaborative Filtering (CF) is a widely adopted approach, but while advanced techniques like graph neural networks (GNNs) and self-supervised learning (SSL) have enhanced CF models for better user representations, they often lack the ability to provide explanations for the recommended items. Explainable recommendations aim to address this gap by offering transparency and insights into the recommendation decision-making process, enhancing users' understanding. This work leverages the language capabilities of Large Language Models (LLMs) to push the boundaries of explainable recommender systems. We introduce a model-agnostic framework called \model, which enables LLMs to provide comprehensive explanations for user behaviors in recommender systems. By integrating collaborative signals and designing a lightweight collaborative adaptor, the framework empowers LLMs to understand complex patterns in user-item interactions and gain a deeper understanding of user preferences. Our extensive experiments demonstrate the effectiveness of \model, showcasing its ability to generate comprehensive and meaningful explanations that outperform baseline approaches in explainable recommender systems. We open-source our model implementation at: \textcolor{blue}{\url{https://github.com/HKUDS/XRec}}.
\end{abstract}

\input{intro}
\input{model}

\input{solution}
\input{eval}

\input{relate}
\input{conclusion}

\clearpage
\input{limitation}
\bibliography{refs}
\appendix
\input{appendix}

\end{document}

%% file: intro.tex
\section{Introduction}
\label{sec:intro}

With the overwhelming abundance of content and products available in online platforms, users frequently encounter the daunting challenge of information overload. In response, recommender systems emerge as indispensable tools that aim to alleviate this burden. These systems effectively filter through the vast array of options and present users with tailored recommendations that are both relevant and personalized, aligning with their unique preferences and interests~\cite{zhang2019deep}.

Among the diverse range of recommendation techniques available, Collaborative Filtering (CF) framework emerges as a prominent and widely embraced approach within the field of recommender systems. CF operates on the fundamental premise that users who have demonstrated similar preferences in the past, such as common item ratings or similar purchase histories, are likely to exhibit comparable preferences when it comes to future recommendations~\cite{chen2021neural}.

In recent years, the field of collaborative filtering algorithms has undergone a remarkable revolution with the emergence of deep learning techniques. This transformative wave has brought about the integration of diverse neural network architectures, including Attention mechanisms \cite{chen2017attentive}, Graph Neural Networks (GNNs) \cite{he2020lightgcn}, and Self-Supervised Learning (SSL) \cite{xia2023automated}. Notably, the incorporation of GNNs in collaborative filtering models has yielded significant advancements by effectively capturing complex relational information and enhancing recommendation performance while preserving high-order collaborative dependencies. Moreover, self-supervised recommender systems have emerged as a promising solution to address the challenge of data sparsity. These systems leverage self-supervised learning signals to augment the available data, aiming to enhance recommendations.

While existing collaborative filtering models excel at providing accurate recommendation results, there remains a critical aspect that has not received adequate attention: understanding the underlying reasons behind observed user-item interactions. Explainable recommendations aim to address this gap by providing transparency to users, offering insights into the decision-making process behind recommendations. This not only enhances users' understanding of their own preferences, but also fosters trust in recommendation algorithms.

Several research studies have dedicated their focus to generating explanations for user-item interactions. Notably, Att2Seq \cite{dong2017learning} and NRT \cite{li2017neural} employ attention mechanisms and recurrent neural networks (RNNs) to generate textual explanations. Recent advancements have further explored the utilization of Transformer \cite{li2021personalized} and GPT2 \cite{li2023personalized} for text generation, providing valuable insights into recommendation results. However, these approaches face a common challenge arising from the limited availability of explanation data, which hinders their ability to generate high-quality explanations. It is also important to emphasize that ID-based methods heavily rely on ID embeddings, resulting in limited generalization capabilities and difficulties when adapting to unseen users and items in a zero-shot recommendation scenario. \\\vspace{-0.12in}

\noindent \textbf{Presented Work}. In light of the recent advancements in Large Language Models (LLMs), our primary objective is to push the boundaries of explainable recommender systems by harnessing the exceptional language capabilities of LLMs. To this end, we introduce model-agnostic \model, a groundbreaking collaborative instruction-tuning framework that enables large language models to provide comprehensive explanations for user behaviors within recommender systems. Within our \model\ framework, we equip LLMs with the unique ability to comprehend the intricate patterns of user-item interactions through the integration of collaborative signals via a collaborative instruction-tuning paradigm. In order to bridge the representation space of collaborative relationships and the language semantic space, we design a lightweight collaborative adaptor that incorporates behavior-aware collaborative signals into the LLMs, facilitating a deeper understanding of user preferences.

We conducted a series of thorough experiments to validate the effectiveness and superior performance of our proposed framework, \model, in generating comprehensive and meaningful explanations within recommender systems. In addition to that, we conducted ablation studies and investigated the robustness of our model, providing further evidence of its effectiveness. 

%% file: model.tex
\section{Preliminaries}
\label{sec:model}

\subsection{Explainable Recommendation}
% Explainable Recommendation is an important component of Recommender Systems (RS), which aims to provide personalized recommendations based on user-item interactions. Different from \textit{interpretable} recommendation, which delineates the influence of each user-item interaction by assigning weights~\cite{ying2019gnnexplainer}, explainable recommendation strives to generate explicit textual explanations, thereby enabling users to understand the reasons behind suggested items. These systems typically utilize techniques such as natural language generation integrated with model predictions~\cite{li2023personalized}. Specifically, for each interaction between user $u$ and item $i$, the generated explanation can be represented as:
Explainable Recommendation is essential in recommender systems as it clarifies the underlying reasons for user-item interactions. Our primary goal is to create clear textual explanations that allow us to understand the rationale behind each recommendation. Specifically, for each interaction between a user $u$ and an item $i$, the explanations generated can be described as follows:
\begin{equation}
    \textit{explanation}(u, i) = \textit{generate}(u, i, \mathcal{X}_u,\mathcal{X}_i,\tau)
\end{equation}
In this context, $\mathcal{X}_u$ and $\mathcal{X}_i$ represent the interaction histories of user $u$ and item $i$, respectively. The symbol $\tau$ denotes any additional side information related to both the users and the items.

% To better refine the accuracy of representations, we incorporate textual information in the form of user and item profiles. Researchers have utilized the reasoning capabilities of LLMs to generate detailed profiles for users and items~\cite{ren2023representation,xi2023towards}, which are designed to offer deeper insights into user preferences and item characteristics.

Building upon recent advancements in text-based profile generation~\cite{ren2023representation,xi2023towards}, we enhance the explanation generation paradigm in recommender systems. Our method involves incorporating textual information into the generation of item profiles, utilizing a predefined item prompt ($\mathcal{P}_I$) and item description ($\mathcal{D}$).
\begin{equation}
    \mathcal{I} = \textit{LLMs}(\mathcal{P}_I, \mathcal{D})
\end{equation}
% For users, rather than directly using descriptions, we enrich the user profiles by incorporating interactions with previously profiled items. This is achieved by sampling a subset of items interacted with by the user, creating a more comprehensive representation of user preferences:
In addition to item descriptions, we extend our approach to user profiles by considering their interactions with previously profiled items. This is achieved by sampling the items that the user has interacted with, resulting in a more comprehensive representation of their preferences and interests.
\begin{equation}
    \mathcal{U} = \textit{LLMs}(\mathcal{P}_U, \{\mathcal{I}_i : i \in \mathcal{N}_u\})
\end{equation}
We denote the set of items interacted with by user $u$ as $\mathcal{N}_u$, and use $\mathcal{P}_U$ as the user profile prompt.

\subsection{Graph Collaborative Filtering}
% To capture collaborative information about user and items, Graph Neural Networks (GNNs) are used for generating user/item representations, in the form of embeddings. This process allows nodes to assimilate information from the entire graph through successive rounds of message passing, culminating in embeddings that encapsulate high-order connectivity characteristics~\cite{wang2019neural}. Specifically, in a graph neural network $\mathcal{G}$ with $L$ layers, node $u$'s $l^{th}$ layer embedding is computed as:
Graph Neural Networks (GNNs) have proven to be effective frameworks for capturing collaborative relationships between users and items, taking into account high-order dependencies. Through multiple rounds of message passing, nodes in the user-item interaction graph assimilate information and generate embeddings that capture these collaborative relationships~\cite{wang2019neural}. To encode the user-item interaction graph $\mathcal{G}$ using $L$ layers of GNNs, the $l^{th}$ layer embedding of a user node $u$ or an item node $i$ is computed as follows:
\begin{equation}
    e_u^{(l)} = \textit{AGG}\left(e_u^{(l-1)}, \{e_i^{(l-1)} \mid i \in \mathcal{N}_u\}\right)
\end{equation}
% where $\mathcal{N}_u$ denotes the neighborhood of node $u$, and $\textit{AGG}$ represents the aggregation function, which varies across different models. Notice that $e_u^{(0)}$ is randomly initialized and is the only trainable layer among all embedding layers. By propagating through $L$ layers, GNNs derive $L+1$ distinct embeddings for each node $(e^{(0)}, e^{(1)}, \dots, e^{(L)})$. These embeddings are subsequently concatenated to form the final node embedding.
In the context of graph collaborative filtering, $\mathcal{N}_u$ refers to the neighborhood of node $u$, while $\textit{AGG}(\cdot)$ is the aggregation function, which can vary across different models. By utilizing $L$ layers of propagation, graph neural networks (GNNs) generate $L+1$ distinct embeddings for each node, namely $(e^{(0)}, e^{(1)}, \dots, e^{(L)})$. These embeddings are then concatenated to form the final node embedding.

The embeddings of users and items are generated with the objective of maximizing the probability, considering the historical user interactions:
% The purpose of generate these embeddings $\mathbf{e}_u, \mathbf{e}_i$ is to maximize the probability:
\begin{equation}
    p(\mathbf{e}|\mathcal{X}) \propto p(\mathcal{X}|\mathbf{e})p(\mathbf{e})
\end{equation}
% for each user in $U = \{u_1, u_2, \dots, u_m\}$ and item in $I = \{i_1, i_2, \dots, i_n\}$, where $\mathcal{X}$ represents user-item historical interactions. The predictive scoring function, represented by the inner product of the user and item embeddings, is defined as:
For each user in the set $U = {u_1, u_2, \dots, u_m}$ and each item in the set $I = {i_1, i_2, \dots, i_n}$, where $\mathcal{X}$ represents the historical interactions between users and items. The predictive scoring function, which is determined by the inner product of the user and item embeddings, can be defined as:
\begin{equation}
    \hat{y}_{ui} = \mathbf{e}_u^T \cdot \mathbf{e}_i
\end{equation}
Assuming normalization, the range of $\hat{y}_{ui}$ is between $0$ and $1$, representing the predicted likelihood of user $u$ interacting with item $i$.

%% file: solution.tex
\section{Methodology}
\label{sec:solution}

\begin{figure*}[t]
    \centering
    \includegraphics[width=0.96\textwidth]{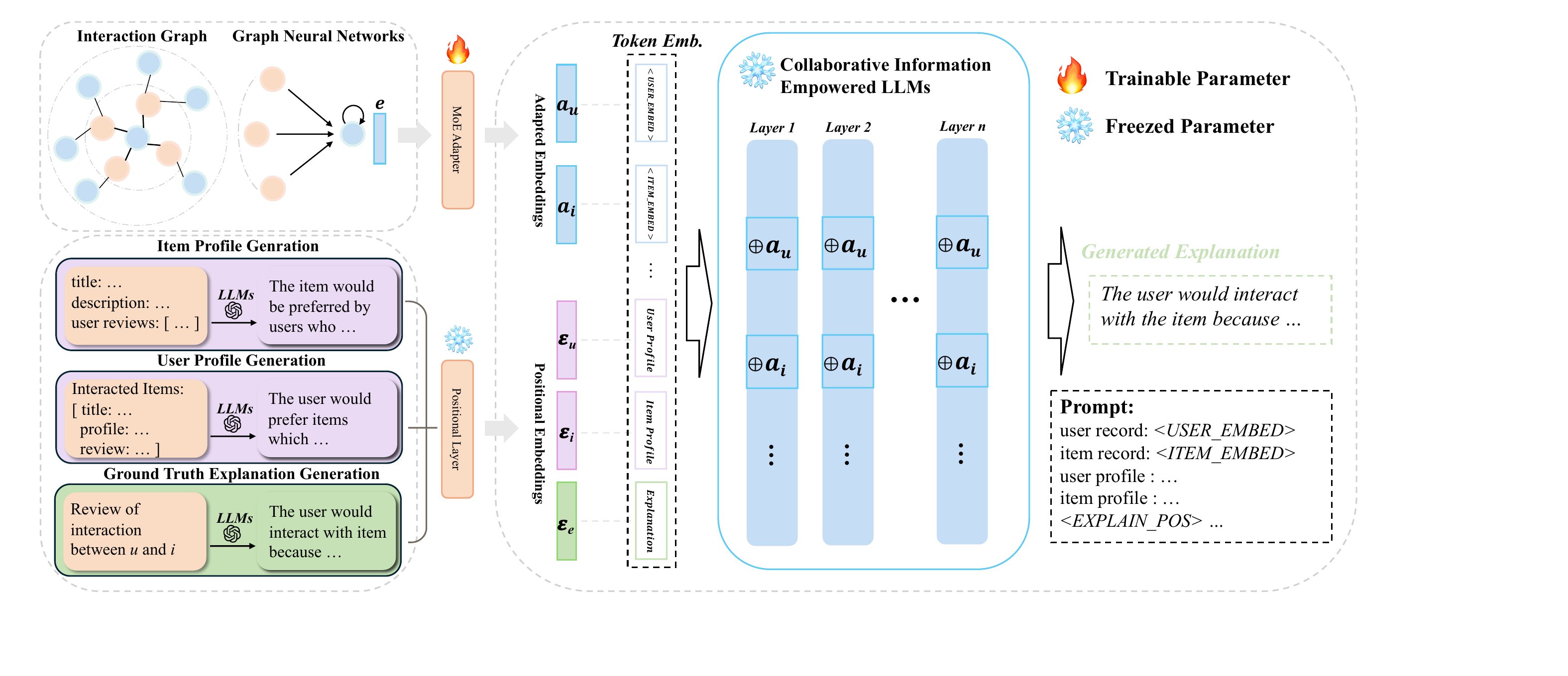}
    \vspace{-0.1in}
    \caption{The overall architecture of our \model. (i). \textbf{Collaborative Relation Tokenizer}: Transforms complex user-item relationships into latent embeddings using GNNs; (ii) \textbf{Collaborative Information Adapter}: A lightweight adapter that integrates collaborative signals into LLMs. (iii) \textbf{Unifying CF with LLM}: Integrates collaborative filtering insights directly into large language models, enabling them to generate insightful explanations.}
    \vspace{-0.2in}
    \label{fig:framework}
\end{figure*}

In this section, we provide a comprehensive overview of our \model, which is specifically designed to generate explanations for user-item interactions. The goal of our model is to uncover the underlying reasons behind these interactions and shed light on the factors influencing user behavior. By unifying graph collaborative filtering and large language models, our \model\ aims to provide insightful explanations which help users understand why certain interactions occur and enhance the transparency of the recommendation process.

\subsection{Collaborative Relation Tokenizer}
To efficiently capture the collaborative relationships between a large number of users and items and reflect their interaction patterns, natural language falls short, but representations offer a powerful alternative. In our \model, we harness the capabilities of graph neural networks as the tokenizer to encode the high-order collaborative relational information into a latent embedding space, enabling effective modeling of complex user preference.

\noindent \textbf{Graph-based Message Passing.} The collaborative graph tokenizer in our approach utilizes message passing mechanisms to propagate and aggregate information across the user-item interaction graph, facilitating the learning of representations for user and item nodes. In our framework, we employ LightGCN~\cite{he2020lightgcn} as the backbone for effective collaborative information aggregation.
\begin{equation}
\begin{aligned}
    \mathbf{e}_u^{(l+1)} = \sum_{i \in \mathcal{N}_u} \frac{1}{\sqrt{|\mathcal{N}_u|} \sqrt{|\mathcal{N}_i|}} \mathbf{e}_i^{(l)}, \\
    \mathbf{e}_i^{(l+1)} = \sum_{u \in \mathcal{N}_i} \frac{1}{\sqrt{|\mathcal{N}_i|} \sqrt{|\mathcal{N}_u|}} \mathbf{e}_u^{(l)}
\end{aligned}
\end{equation}
The user and item final embeddings are computed by averaging all layer embeddings.
\begin{equation}
    \mathbf{e}_u = \sum_{k=0}^K \frac{1}{K+1} \mathbf{e}_u^{(k)}, \quad \mathbf{e}_i = \sum_{k=0}^K \frac{1}{K+1} \mathbf{e}_i^{(k)}
\end{equation}

\noindent\textbf{Tokenizer Optimization with CF signals.}
To optimize our collaborative graph tokenizer using implicit feedback signals from user interactions, we utilize the Bayesian Personalized Ranking (BPR) loss as a supervision signal to guide the generation of user and item embeddings, which is defined as:
\begin{equation}
 L_{BPR} = - \sum_{u=1}^{m} \sum_{i \in \mathcal{N}_u} \sum_{j \notin \mathcal{N}_u} \ln \sigma (\hat{y}_{u,i} - \hat{y}_{u,j})
\end{equation}
$\hat{y}_{u,i}$ denotes the prediction score (inner product) between user $u$ and item $i$, and $\sigma$ denotes the sigmoid function. Additionally, we include a regularization loss to maintain the norm of the embeddings:
\begin{equation}
    L_{reg} = \lambda (\|\mathbf{e}_u^{(0)}\|^2 + \|\mathbf{e}_i^{(0)}\|^2)
\end{equation}
By combining these terms, our joint optimization loss function is formulated as:
\begin{equation}
    L = L_{BPR} + L_{reg}
\end{equation}

\subsection{Collaborative Instruction Tuning for Large Language Models (LLMs)}
To enable LLMs to understand collaborative information among users and items, we introduce a collaborative instruction tuning paradigm. This approach aligns behavior-level information with language-level semantics, thereby incorporating user preferences into the knowledge within LLMs.

\subsubsection{Collaborative Information Adapter}
Given the potentially divergent semantic representation spaces between the behavior-level collaborative information and the textual semantics associated with users and items, our \model\ is equipped with a lightweight yet effective adapter. This adapter serves to align these different modalities, enabling our model to effectively leverage both the collaborative signals and the textual semantics.

To bridge the semantic gap between the input of large language models (LLMs) and our behavior-aware collaborative relation tokens, and to enhance the model's generalization capabilities, we apply a Mixture of Experts (MoE) approach~\cite{hou2022towards} for embedding space adaptation. In this Mixture of Experts architecture, each expert is represented by a linear layer that captures different semantic dimensions, and these experts are then integrated using a learnable gating router mechanism. This allows the model to adaptively combine the different semantic representations encoded by the various experts, effectively bridging the gap between the behavior-aware collaborative relation tokens and textual language tokens.

\subsubsection{Unifying CF with LLM}
With the newly adapted embeddings, we are now ready to infuse collaborative information into LLMs. We introduce special tokens to reserve space in the input prompt, and after transforming the prompt into token embeddings, we inject the adapted embeddings into these reserved positions.

However, a challenge arises as each node embedding is represented by only a single token in the input prompt. As the input length increases, the attention weight allocated to each embedding token inevitably diminishes, leading to a potential loss of collaborative information. To address this dilution of influence, we take inspiration from~\cite{qin2023disentangled} and extend the injection of adapted embeddings beyond the initial input prompt. Specifically, we incorporate them into every layer of the LLM at reserved positions. To facilitate this, we modify the key, query, and value projection functions of every layer within the LLMs as follows:
\begin{equation}
    f_{\{q,k,v\}}(\mathbf{x}_i) \quad{+=} \quad \mathbf{W}_{\{q,k,v\}} \cdot \mathbf{a}_i
\end{equation}
Let's denote the projection matrices for the query, key, and value as $\mathbf{W}_{{q,k,v}}$, and $\mathbf{a}_i$ as the adapted embedding. Our approach ensures that the large language models (LLMs) continuously access and integrate the collaborative information throughout their entire structure, not just at the input stage. By injecting the graph-based knowledge into all layers of the LLMs, we not only maintain a robust representation of the collaborative context, but also enable more effective gradient flow directly back to the Mixture of Experts (MoE) module. This innovative integration of language modeling and graph representation learning allows our model to leverage the deep contextual insights provided by the LLMs, while benefiting from the structural patterns recognized by the Graph Neural Networks. \\\vspace{-0.12in}

\begin{figure}[t]
    \centering
    \includegraphics[width=\columnwidth]{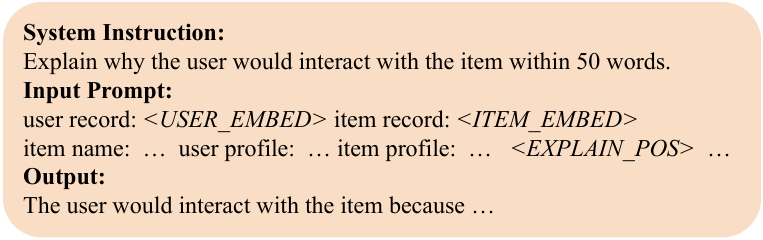}
    \caption{A depiction of model prompt instruction.}
    \vspace{-0.2in}
    \label{fig:prompt}
\end{figure}

\noindent\textbf{Structured Prompt Embedding}.
We employ a structured prompt, illustrated in Figure \ref{fig:prompt}, which integrates various data elements. This process involves tokenizing the prompt and converting it into an embedding space representation. To ensure the special tokens within the prompt are recognized as unique entities, we incorporate them into the tokenizer of the LLM. These specialized tokens are then replaced by their corresponding adapted embeddings in the transformed token embeddings.

Specifically, we define the input prompt as $\mathcal{P}=[p_1, \dots, p_u, \dots, p_i, \dots, p_e, \dots, p_l]$, where each element $p$ represents one input token. The tokens $p_u, p_i, p_e$ denote \textit{<USER\_EMBED>}, \textit{<ITEM\_EMBED>}, \textit{<EXPLAIN\_POS>} respectively. After processing through the positional embedding layer, we denote the output as $\mathcal{E}=[\epsilon_1, \dots, \epsilon_u, \dots, \epsilon_i, \dots, \epsilon_e, \dots, \epsilon_l]$, where each $\epsilon$ is the token embedding of its corresponding token. Subsequently, $\epsilon_u$ and $\epsilon_i$ are replaced by the adapted embeddings $a_u$ and $a_i$, to form the final embedding layer $\mathcal{E'}=[\epsilon_1, \dots, a_u, \dots, a_i, \dots, \epsilon_e, \dots, \epsilon_l]$, which is further used as input of LLMs.

To improve the ability of large language models (LLMs) to generate contextually and syntactically coherent explanations, we aim to minimize the loss between the predicted probabilities of next tokens and the actual next tokens in the sequences. We utilize the negative log-likelihood (NLL) as our training loss, calculated as follows:
\begin{equation}
    \mathcal{L} = -\frac{1}{N} \sum_{i=1}^{N} \sum_{c=1}^{C_i} y_{ic} \cdot \log(\hat{y}_{ic})
\end{equation}
Here, $N$ is the number of explanations, $C_i$ is the character count in each explanation, and $y_{i,c}$ and $\hat{y}_{i,c}$ represent the actual and predicted tokens, respectively. To minimize training complexity, we freeze all parameters within the LLMs, excluding any interactions with the GNN training procedure. The only trainable parameters are those within the Mixture-of-Experts (MoE) model.

\subsubsection{Ground Truth Explanation Generation}
Prior research has directly used user reviews as ground truth explanations for recommender systems~\cite{li2021personalized}. However, these reviews tend to be subjective and may only implicitly convey the user's intentions or sentiments. To address this limitation and improve the quality of ground truth explanations, the application of Large Language Models (LLMs) has been proposed to distill explicit user intentions from their raw reviews.
\begin{equation}
    \textit{explanation}(u, i) = \textit{LLMs}(\mathcal{P}, r_{u, i})
\end{equation}
where $r_{u,i}$ is the review of item $i$ given by user $u$. An example case is shown in the Appendix.

%% file: eval.tex
\section{Evaluation}
\label{sec:eval}

\begin{table*}[t]
    \centering
    \caption{Overall Comparison in Terms of Explainability and Stability. The superscripts \textit{P}, \textit{R}, and \textit{F1} indicate Precision, Recall, and F1-Score, respectively. The subscript std denotes the standard deviation of each score. The best performances are highlighted in bold, and the second-best are underlined.}
    \vspace{-0.1in}
    \resizebox{\textwidth}{!}{
        \begin{tabular}{r|lllllll|llllll}
    		\toprule
    		\multirow{2}{2cm}{\textbf{Metrics}}& \multicolumn{7}{c|}{\textbf{Explainability $\uparrow$}} & \multicolumn{6}{c}{\textbf{Stability $\downarrow$}} \\ \cline{2-14}
            & GPTScore & BERTScore$^P$ & BERTScore$^R$ & BERTScore$^{F1}$ & BARTScore & BLEURT & USR & GPT$_{std}$ & BERT$^{P}_{std}$ & BERT$^{R}_{std}$ & BERT$^{F1}_{std}$ & BART$_{std}$ & BLEURT$_{std}$\\ \midrule
            \multicolumn{14}{c}{Amazon-books}\\ \midrule
            \multicolumn{1}{l|}{Att2Seq} & 76.08 & 0.3746 & 0.3624 & 0.3687 & -3.9440 & -0.3302 & 0.7757 & 12.56 & 0.1691 & 0.1051 & 0.1275 & 0.5080 & 0.299\\
            \multicolumn{1}{l|}{NRT} & 75.63 & 0.3444 & 0.3440 & 0.3443 & -3.9806 & -0.4073 & 0.5413 & 12.82 & 0.1804 & 0.1035 & 0.1321 & 0.5101 & 0.3104\\ 
            \multicolumn{1}{l|}{PETER} & 77.65 & \textbf{0.4279} & 0.3799 & 0.4043 & -3.8968 & -0.2937 & 0.8480 & 11.21 & 0.1334 & 0.1035 & 0.1098 & 0.5144 & 0.2667\\ 
            \multicolumn{1}{l|}{PEPLER} & 78.77 & 0.3506 & 0.3569 & 0.3543 & -3.9142 & -0.2950 & 0.9563 & 11.38 & 0.1105 & \underline{0.0935} & 0.0893 & 0.5064 & 0.2195\\
            \multicolumn{1}{l|}{Ours (w/o profile)} & \underline{81.77} & \underline{0.4194} & \underline{0.4004} & 0.4106 & \underline{-3.8218} & \underline{-0.1294} & \underline{\textbf{1.0000}} & \underline{\textbf{9.60}} & \textbf{0.0819} & 0.0955 & \textbf{0.0786} & \textbf{0.4799} & \underline{0.1803}\\
            \multicolumn{1}{l|}{Ours} & \textbf{82.57} & 0.4193 & \textbf{0.4038} & \textbf{0.4122} & \textbf{-3.8035} & \textbf{-0.1061} & \underline{\textbf{1.0000}} & \underline{\textbf{9.60}} & \underline{0.0836} & \textbf{0.0920} & \underline{0.0800} & \underline{0.4832} & \textbf{0.1780}\\
            \midrule
            \multicolumn{14}{c}{Yelp}\\ \midrule
            \multicolumn{1}{l|}{Att2Seq} & 63.91 & 0.2099 & 0.2658 & 0.2379 & -4.5316 & -0.6707 & 0.7583 & 15.62 & 0.1583 & 0.1074 & 0.1147 & \underline{0.5616} & 0.247\\ 
            \multicolumn{1}{l|}{NRT} & 61.94 & 0.0795 & 0.2225 & 0.1495 & -4.6142 & -0.7913 & 0.2677 & 16.81 & 0.2293 & 0.1134 & 0.1581 & \textbf{0.5612} & 0.2728 \\ 
            \multicolumn{1}{l|}{PETER} & 67.00 & 0.2102 & 0.2983 & 0.2513 & -4.4100 & -0.5816 & 0.8750 & 15.57 & 0.3315 & 0.1298 & 0.2230 & 0.5800 & 0.3555\\ 
            \multicolumn{1}{l|}{PEPLER} & 67.54 & 0.2920 & 0.3183 & 0.3052 & -4.4563 & -0.3354 & 0.9143 & 14.18 & 0.1476 & \textbf{0.1044} & 0.1050 & 0.5777 & 0.2524\\
            \multicolumn{1}{l|}{Ours (w/o profile)} & \underline{71.81} & \underline{0.3879} & \underline{0.3427} & \underline{0.3657} & \underline{-4.4035} & \underline{-0.2486} & \underline{\textbf{1.0000}} & \underline{12.71} & \underline{0.1087} & 0.1072 & \underline{0.0919} & 0.5717 & \textbf{0.2272} \\
            \multicolumn{1}{l|}{Ours} & \textbf{74.53} & \textbf{0.3946} & \textbf{0.3506} & \textbf{0.3730} & \textbf{-4.3911} & \textbf{-0.2287} & \underline{\textbf{1.0000}} & \textbf{11.45} & \textbf{0.0969} & \underline{0.1048} & \textbf{0.0852} & 0.5770 & \underline{0.2322}\\
            \midrule
            \multicolumn{14}{c}{Google-reviews}\\ \midrule
            \multicolumn{1}{l|}{Att2Seq} & 61.31 & 0.3619 & 0.3653 & 0.3636 & -4.2627 & -0.4671 & 0.5070 & 17.47 & 0.1855 & 0.1247 & 0.1403 & 0.6663 & 0.3198 \\ 
            \multicolumn{1}{l|}{NRT} & 58.27 & 0.3509 & 0.3495 & 0.3496 & -4.2915 & -0.4838 & 0.2533 & 19.16 & 0.2176 & 0.1267 & 0.1571 & 0.6620 & 0.3118 \\ 
            \multicolumn{1}{l|}{PETER} & 65.16 & 0.3892 & 0.3905 & 0.3881 & \underline{-4.1527} & -0.3375 & 0.4757 & 17.00 & 0.2819 & 0.1356 & 0.2005 & 0.6701 & 0.3272 \\ 
            \multicolumn{1}{l|}{PEPLER} & 61.58 & 0.3373 & 0.3711 & 0.3546 & -4.1744 & -0.2892 & 0.8660 & 17.17 & \underline{0.1134} & \textbf{0.1161} & \underline{0.0999} & 0.6752 & 0.2484\\
            \multicolumn{1}{l|}{Ours (w/o profile)} & \textbf{69.71} & \underline{0.4427} & \textbf{0.4187} & \underline{0.4310} & \textbf{-4.1142} & \textbf{-0.2026} & \textbf{0.9997} & \textbf{14.09} & 0.1180 & 0.1171 & 0.1034 & \textbf{0.6465} & \textbf{0.2439}\\
            \multicolumn{1}{l|}{Ours} & \underline{69.12} & \textbf{0.4546} & \underline{0.4069} & \textbf{0.4311} & -4.1647 & \underline{-0.2437} & \underline{0.9993} & \underline{14.24} & \textbf{0.0972} & \underline{0.1163} & \textbf{0.0938} & \underline{0.6591} & \underline{0.2452}\\
            \bottomrule
        \end{tabular}
    }
    \label{tab:Genration Performance}
    \vspace{-0.15in}
\end{table*}

\subsection{Experimental Settings}
\noindent\textbf{Datasets.}
To evaluate \model, we utilize three prominent public datasets that offer distinct perspectives on user-item interactions: \textbf{Amazon} \cite{ni2019justifying}: This dataset aggregates the purchasing behaviors of users within Amazon's books category. It includes not only user ratings but also the textual reviews they provide after making a purchase. \textbf{Yelp}: This dataset captures the interactions between users and businesses, with a focus on the service industry. It contains both user ratings and reviews. \textbf{Google} \cite{li2022uctopic, yan2023personalized}: Centered on user interactions recorded through Google Maps, this dataset incorporates the metadata of businesses as well as the feedback provided by users. Detailed statistics of these datasets can be found in Table~\ref{tab:data statistics}. \\\vspace{-0.15in}

\begin{table}[t]
    \centering
    \small
    % \footnotesize
    % \scriptsize
    \caption{Statistics of the experimental datasets.}
    \vspace{-0.15in}
    \begin{tabular}{cccc}
        \toprule
        Dataset & \#Users & \#Items & \#Interactions \\
        \midrule
        Amazon & 15,349 & 15,247 & 360,839 \\
        Yelp & 15,942 & 14,085 & 393,680 \\
        Google & 22,582 & 16,557 & 411,840 \\
        \bottomrule
        %\hline
    \end{tabular}
    \label{tab:data statistics}
    \vspace{-0.15in}
\end{table}

\noindent\textbf{Evaluation Metrics.}
When evaluating our \model, we employ a suite of metrics designed to capture the semantic explainability and stability of the generated explanations. Traditional n-gram based metrics like BLEU and ROUGE prove unsuitable, as they fail to grasp the underlying semantic meaning. For instance, "the weather is cold" and "it's freezing" convey the same meaning, yet would score poorly due to a lack of n-gram overlap.

Instead, we employ advanced metrics that incorporate semantic understanding: GPTScore \cite{wang2023chatgpt} aligns with human judgment by comparing the semantic similarity between generated and ground truth explanations. BERTScore \cite{zhang2019bertscore} utilizes contextual embeddings from BERT to compute token-level cosine similarity. BARTScore \cite{yuan2021bartscore} leverages the BART model, conceptualizing evaluation as a text generation task that assigns scores based on the probability of regenerating reference texts. BLEURT \cite{sellam2020bleurt} employs a novel pre-training approach with synthetic data to enhance generalization. USR \cite{li2021personalized} (Unique Sentence Ratio) measures the uniqueness of generated explanations by calculating the ratio of unique to total sentences.

To further assess quality stability, we analyze the standard deviations of these scores, with lower values indicating more consistent performance. The overall results are shown in Table~\ref{tab:Genration Performance}.

\noindent\textbf{Compared Methods.}
We compare our model's performance against the following baselines:\vspace{-0.05in}
\begin{itemize}[leftmargin=*]

    \item \textbf{Att2Seq}~\cite{dong2017learning}: Utilizes an attention-based attribute-to-sequence model to generate reviews based on attribute information.
    
    \item \textbf{NRT}~\cite{li2017neural}: Predicts ratings and generates abstractive tips for recommendations using multi-task learning to optimize parameters.
    
    \item \textbf{PETER}~\cite{li2021personalized}: PETER is a personalized transformer model for explainable recommendation. It maps user and item IDs to the generated explanation text, connecting the IDs and words. Since the datasets lack word features, PETER is used instead of PETER+, which incorporates additional features.

    \item \textbf{PEPLER}~\cite{li2023personalized}: PEPLER leverages pretrained transformer to generate explainable recommendations based on prompts that incorporate user and item ID vectors. To bridge the gap between these prompts and the pretrained model, the approach proposes sequential tuning and recommendation as regularization strategies. There are several variants of the PEPLER, and in this case, the researchers have opted to use the continuous prompt learning version.
    
    \item \textbf{Ours (w/o profile)}: For a fair comparison across all the models, we remove the user and item profiles from the input to our model, as the other baselines do not have access to this information.
\end{itemize}

\noindent\textbf{Implementation Details.}
When generating the graph embeddings, we configure the embedding dimension to 64 and use a batch size of 1024. To optimize the training process, we implement an early stopping mechanism based on Recall@20, allowing for up to 10 patience steps. In our Mixture of Experts (MoE) setup, we utilize 8 experts and incorporate a dropout rate of 0.2, with an additional noise factor of 0.01 at the gating router. The LLM we use is built upon the LLaMA2-7B architecture. Additionally, we employ the gpt-3.5-turbo model for generating datasets and for computing the GPTScore. For the explanations, we ensure they are no longer than 50 words for both the ground truth and the generated explanations.

\subsection{Performance Comparison}
To demonstrate the superiority of our model in explainability and stability, we conduct comparative analyses against four baseline methods across three datasets. The results are summarized in Table~\ref{tab:Genration Performance} and reveal several key findings:
\begin{itemize}[leftmargin=*]
    \item Our model consistently outperforms the baselines. Even after removing user and item profiles, the Ours (w/o profile) variant still demonstrates significant superiority. This success can be attributed to two key factors: i) Our model effectively captures collaborative information, enhancing the representation of rich semantics from user interaction behaviors, going beyond just textual information. ii) The model achieves strong alignment between the behavior-level collaborative information and text-level semantics.
    
    \item A standout feature of our model is its Unique Sentence Ratio (USR), which is nearly 1. This indicates that XRec generates truly unique explanations for each distinct user-item interaction. This remarkable level of uniqueness in the generated explanations represents a significant breakthrough. No previous work has achieved such a high degree of personalization in model outputs.
\end{itemize}

Our model enhances learning efficacy and boosts overall performance by strategically exploiting the synergistic strengths across domains.

\subsection{Ablation Study}

\begin{figure}[t]
    \centering
    \includegraphics[width=\columnwidth]{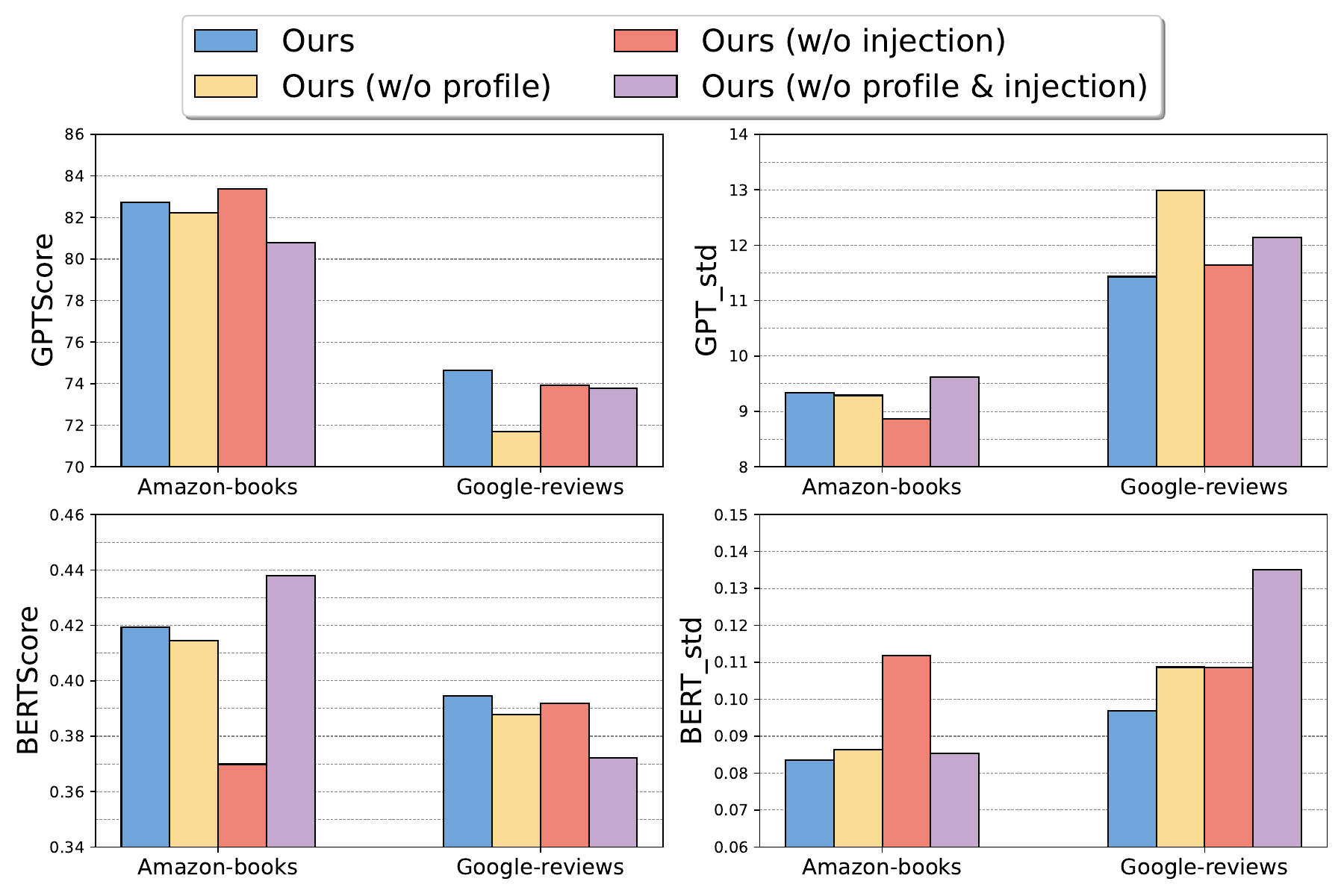}
    \vspace{-0.25in}
    \caption{Ablation Study on Variant Models: Higher scores in GPTScore and BERTScore suggest improved explainability, while lower scores in GPT\_std and BERT\_std indicate enhanced stability.}
    \label{fig:ablation}
    \vspace{-0.2in}
\end{figure}

In this section, we conduct ablation studies to explore the impact of two pivotal components in our model: user/item profiles and the injection of collaborative information. We compare four model variants - i) our complete model with all features (\textbf{Ours}); ii) \textbf{Ours (w/o profile)} which omits user and item profiles; iii) \textbf{Ours (w/o injection)} which retains aligned embeddings in the prompts but does not inject them into the LLM layers, and iv) \textbf{Ours (w/o profile \& injection)} lacking both profiles and embedding injection. To rigorously assess explainability and stability, we evaluate these variants using GPTScore and BERTScore on the Amazon-books and Google-reviews datasets, including their standard deviations, which sheds light on the critical role each of these elements plays in driving the model's performance and capabilities.

The results in Figure~\ref{fig:ablation} show our complete model (Ours) outperforms other variants in explainability and stability, highlighting its superior capability. i) While \textbf{Ours (w/o profile)} declines only slightly, it exhibits a significant reduction in stability compared to Ours, underscoring the critical importance of user/item profiles for explanation stability. ii) Similarly, \textbf{Ours (w/o injection)} reports much lower scores, emphasizing the value of incorporating neighborhood knowledge to improve performance. iii) Notably, \textbf{Ours (w/o profile \& injection)} exhibits the lowest scores, confirming the synergistic combination of user/item profiles and knowledge injection is crucial for optimal performance. These findings underscore the complementary and synergistic contribution of these two critical components in driving the superior capabilities of our complete model. 

\begin{figure}[t]
    \centering
    \includegraphics[width=\columnwidth]{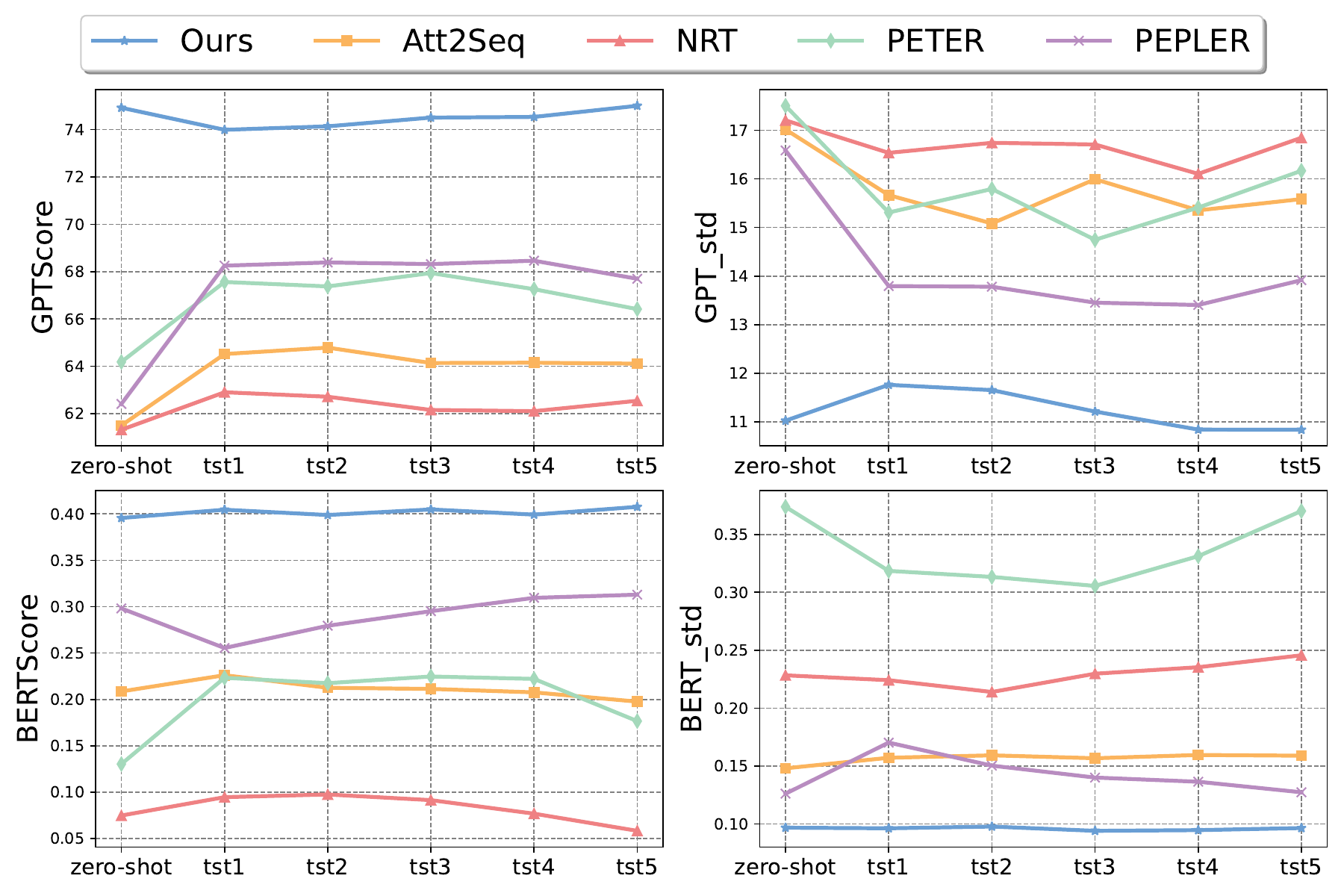}
    \vspace{-0.25in}
    \caption{Experiments of different data sparsity.}
    \label{fig:Analysis}
    \vspace{-0.2in}
\end{figure}

\subsection{Model Robustness against Data Sparsity}
To evaluate our model's generalization capabilities, we conducted experiments across datasets with varying data sparsity. We segmented the testing data into five subsets (tst1 to tst5) based on the frequency of user appearances in the training data. This allowed us to systematically examine the model's effectiveness across a spectrum of user familiarity, from rare to frequent users. Additionally, we introduced a zero-shot testing dataset consisting solely of users not encountered during training, which tested the model's ability to address the cold-start problem. The evaluation results, summarized in Figure 1, highlight several key findings:
\begin{itemize}[leftmargin=*]
    \item The model demonstrates robust performance across all subsets, with noticeably better results as user frequency decreases. This trend suggests our model effectively leverages collaborative information, even with limited user interactions.

    \item In the zero-shot scenario, lacking any prior user data, our model not only outperforms baselines but also performs comparably to the other subsets, from tst1 to tst5. This capability is valuable for new user recommendations, highlighting the practical utility of our approach in real-world applications with incomplete user data.
\end{itemize}

These findings underscore the efficacy of our model in scenarios that traditionally challenge recommender systems, such as those involving new or infrequent users. The model's success in the zero-shot learning confirms its robust generalization capabilities and highlights its potential to mitigate the cold-start problem, where new users or items lack historical interaction data. By maintaining high levels of explainability and stability across diverse scenarios, the model proves its suitability for deployment in dynamic environments where user behaviors and item catalogs frequently change.

%% file: relate.tex
\section{Related Work}
\label{sec:relate}

\subsection{Explainable Recommendation}
Explainable recommender systems have attracted considerable attention due to their ability to enhance user satisfaction and provide transparency in recommendation processes. Early approaches primarily relied on extracting attributes from users and items' side information to generate explanations, employing techniques such as attention mechanisms \cite{dong2017learning} and recurrent neural networks (RNNs) \cite{li2017neural}. In recent studies, researchers have further explored the application of advanced models like Transformer \cite{li2021personalized} and GPT2 \cite{li2023personalized} for offering explanations regarding user behaviors in recommender systems.

However, a predominant issue with most existing solutions for explainable recommendations is their heavy reliance on ID-based approaches. This dependency significantly restricts their generalization ability, especially when confronted with challenges such as data sparsity and zero-shot recommendation scenarios. Furthermore, the scarcity of explanatory data presents additional obstacles, as it poses challenges for existing methods to deliver explanations of high quality and comprehensiveness. Given the aforementioned obstacles, we propose the development of a novel large language model as an explainer for recommender systems. This model not only uncovers the underlying reasons behind user-item interactions, but also demonstrates robust generalization capabilities, even in zero-shot recommendation scenarios.

\subsection{GNNs for Recommendation}
Graph Neural Networks (GNNs) have become a core part of improving collaborative filtering models. They offer an effective way to capture the complex high-level interactions in recommendation systems \cite{ying2018graph,ren2023disentangled}. These networks leverage the natural relational structure of data, enabling a sophisticated understanding of the intricate dependencies that define user-item interactions. Recommender systems like LightGCN \cite{he2020lightgcn} and Star-GCN \cite{zhang2019star} have set the standard, using iterative message passing to model and enhance collaborative relationships. To address the challenge of data sparsity, researchers have integrated self-supervised learning with the graph-based collaborative filtering approach. This introduces new methods to enrich the learning process and improve recommendation quality \cite{yang2023generative,yang2023knowledge,yao2021self}.

Drawing inspiration from the aforementioned research endeavors that emphasize GNN-enhanced recommender systems, we have successfully developed our advanced language model, \model. By incorporating GNN as the collaborative relation encoders, our model excels at capturing intricate user dependencies at higher orders. Through our collaborative instruction-tuning framework, we equip LLMs with the ability to recognize and leverage collaborative signals among users, effectively aligning behavior-level user preferences with the language semantic space. This enables our model to provide comprehensive textual explanations that correspond to user interaction behaviors.

%% file: conclusion.tex
\section{Conclusion}
\label{sec:conclusion}
This work presents a novel framework, \model, that seamlessly integrates the graph-based collaborative filtering paradigm with the capabilities of Large Language Models (LLMs) to generate comprehensive and insightful explanations for recommendation outputs. By leveraging the inherent collaborative relationships encoded within the user-item interaction graph, \model\ is able to effectively capture the high-order dependencies that underlie user preferences and item associations. \model\ introduces a specialized collaborative information adaptor, which serves as the critical bridge for establishing a strong connection between the collaborative signals and the rich textual semantics encoded within the LLM. Through extensive experiments, the study's findings underscore the significant advantages of the \model\ framework. Not only does it enhance the explainability of the recommendation process, but it also ensures robustness, particularly in challenging zero-shot scenarios where the framework demonstrates strong generalization capabilities across unseen users and items.

%% file: limitation.tex
\section{Limitation}
\label{sec:limit}

While \model\ demonstrates promising advancements in explainable recommender systems, it also exhibits limitations in terms of data modality diversity. Currently, our approach is constrained to textual and graph-based data, excluding visual inputs such as images and videos. These visual modalities can provide extensive contextual information. For instance, images and videos can reveal aesthetic preferences, cultural trends, and emotional responses, capturing subtle cues that textual data might miss. They also help identify visual patterns that correspond to user behavior and preferences, such as color schemes, styles, and settings, which are crucial for personalization. Recognizing these limitations, future work could explore the integration of multimodal data processing techniques. This approach may potentially enhance the system's predictive accuracy and improve its ability to personalize recommendation explanations, by incorporating advanced image and video analysis to better understand and respond to user preferences.

%% file: appendix.tex
\section{Appendix}
\label{sec:appendix}
The supplementary materials accompanying this work offer a comprehensive and meticulous exploration of the methodologies and techniques utilized in crafting the ground truth explanations and developing the detailed user/item profiles.

\subsection{Generating Ground Truth Explanations}
\begin{figure}[t]
    \centering
    \includegraphics[width=1.0\linewidth]{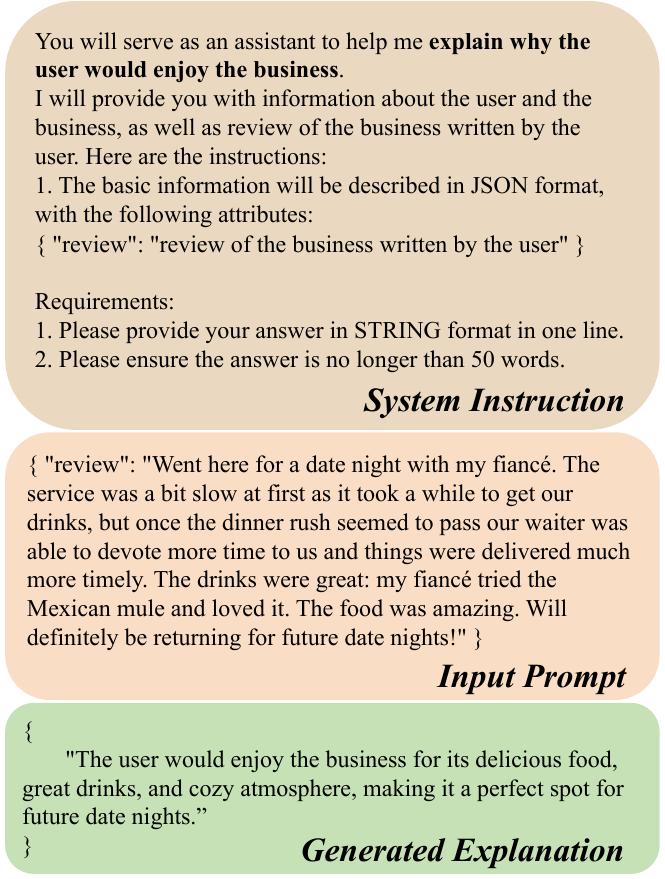}
    \vspace{-0.1in}
    \caption{Case study on the generation of ground truth explanations for recommender systems on Yelp dataset.}
    \label{fig:explanation}
\end{figure}

Figure~\ref{fig:explanation} provides an illustrative example of the ground truth explanation generation process as applied to the Yelp dataset. To maintain a consistent approach across diverse user-item interactions, the instructions given to the language models remain uniform, guiding them to extract the most relevant information that accurately reflects the user's underlying intentions. Notably, the prompt used in this process consists solely of the user's review text, deliberately omitting any additional data about the user or the item being reviewed. This strategic decision helps to minimize the influence of extraneous information, allowing the language model to focus exclusively on discerning and articulating the implicit intentions behind the user's interaction.

\subsection{Item Profile Generation}
\begin{figure*}[t]
    \centering
    \includegraphics[width=0.9\textwidth]{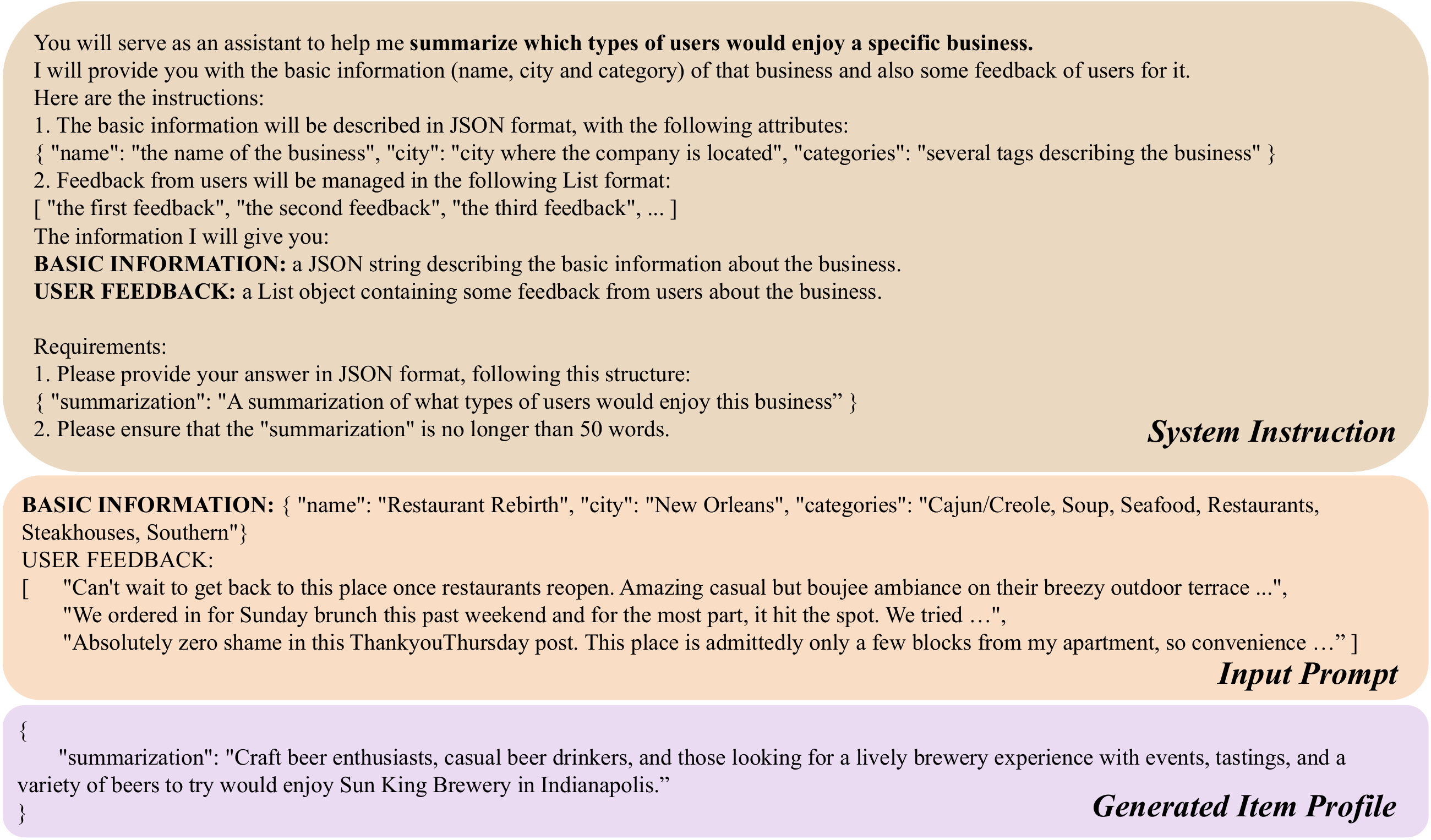}
    \vspace{-0.1in}
    \caption{Case study of item profile generation on Yelp dataset.}
    \label{fig:item profile}
\end{figure*}

Figure~\ref{fig:item profile} illustrates the item profile generation process using the Yelp data. LLMs are fed metadata about the item, such as its name, location, and category, as well as user reviews. This approach allows the LLMs to gain a deeper understanding of the types of users who favor the business. By processing this multifaceted information, the LLMs generate comprehensive profiles that summarize the key characteristics of users inclined towards the item. For instance, the case study profiles capture insights into the preferences of users drawn to beer-related businesses. This granular understanding represents a significant advancement in modeling user-item interactions and preferences.

\subsection{User Profile Generation}
\begin{figure*}[t]
    \centering
    \includegraphics[width=0.9\textwidth]{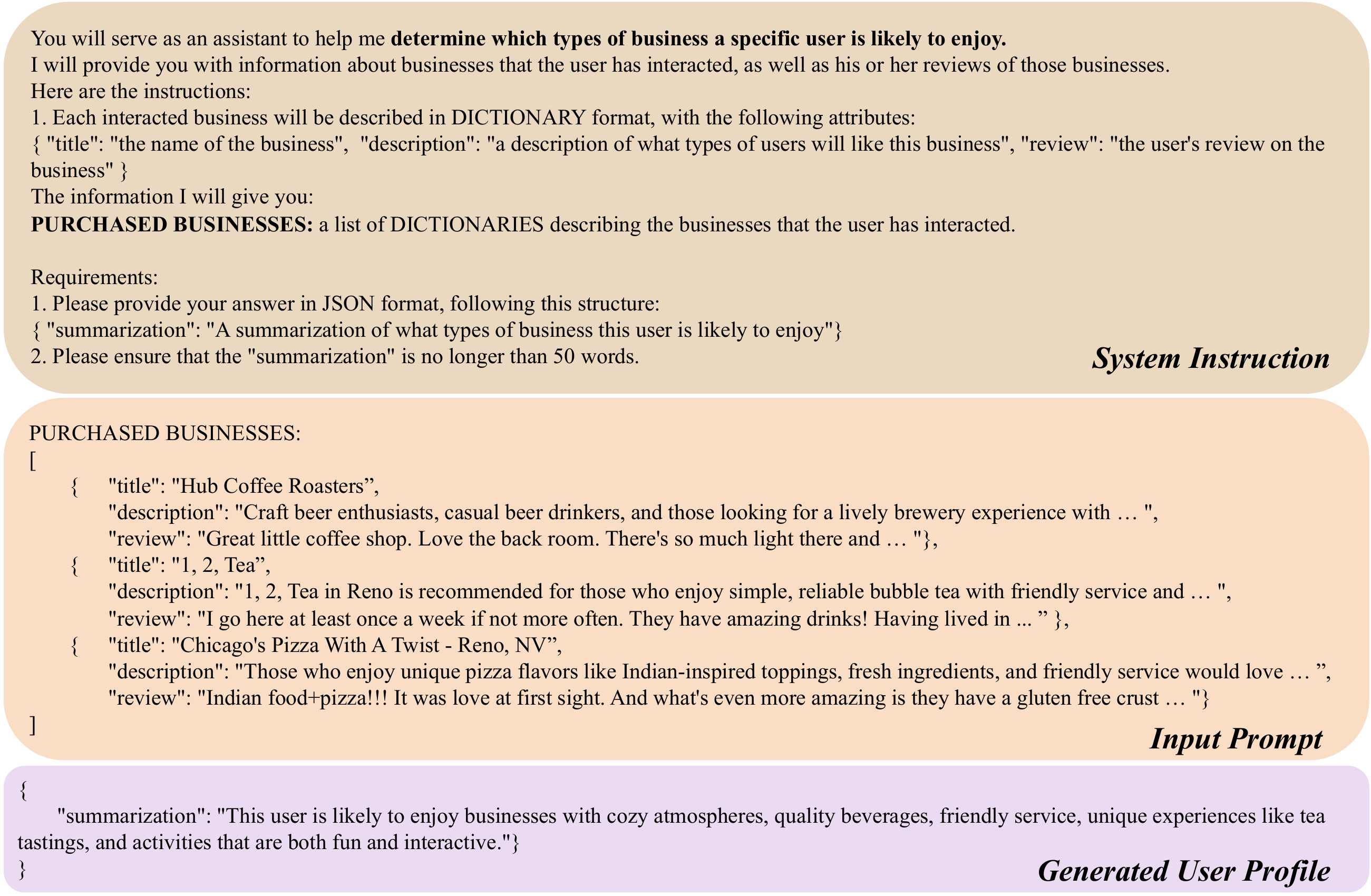}
    \vspace{-0.1in}
    \caption{Case study of user profile generation on Yelp dataset.}
    \vspace{-0.2in}
    \label{fig:user profile}
\end{figure*}

Figure~\ref{fig:user profile} explores user profile generation on the Yelp dataset. Unlike items, users often lack extensive metadata, presenting a unique challenge. To deduce user preferences, the approach relies on Large Language Models (LLMs) to analyze the user's historical interactions. Crucially, the LLMs leverage item descriptions derived from previously generated profiles, refined to better characterize the items. This methodology, incorporating metadata like item titles and reviews, empowers the LLMs to identify the specific types of items a user prefers. By leveraging this multifaceted information, the system develops nuanced user profiles to enable personalized recommendations.

%% file: main.bbl
\begin{thebibliography}{27}
\expandafter\ifx\csname natexlab\endcsname\relax\def\natexlab#1{#1}\fi

\bibitem[{Chen et~al.(2021)Chen, Shi, Li, and Zhang}]{chen2021neural}
Hanxiong Chen, Shaoyun Shi, Yunqi Li, and Yongfeng Zhang. 2021.
\newblock Neural collaborative reasoning.
\newblock In \emph{WWW}, pages 1516--1527.

\bibitem[{Chen et~al.(2017)Chen, Zhang, He, Nie, Liu, and Chua}]{chen2017attentive}
Jingyuan Chen, Hanwang Zhang, Xiangnan He, Liqiang Nie, Wei Liu, and Tat-Seng Chua. 2017.
\newblock Attentive collaborative filtering: Multimedia recommendation with item-and component-level attention.
\newblock In \emph{SIGIR}, pages 335--344.

\bibitem[{Dong et~al.(2017)Dong, Huang, Wei, Lapata, Zhou, and Xu}]{dong2017learning}
Li~Dong, Shaohan Huang, Furu Wei, Mirella Lapata, Ming Zhou, and Ke~Xu. 2017.
\newblock Learning to generate product reviews from attributes.
\newblock In \emph{EACL}, pages 623--632.

\bibitem[{He et~al.(2020)He, Deng, Wang, Li, Zhang, and Wang}]{he2020lightgcn}
Xiangnan He, Kuan Deng, Xiang Wang, Yan Li, Yongdong Zhang, and Meng Wang. 2020.
\newblock Lightgcn: Simplifying and powering graph convolution network for recommendation.
\newblock In \emph{SIGIR}, pages 639--648.

\bibitem[{Hou et~al.(2022)Hou, Mu, Zhao, Li, Ding, and Wen}]{hou2022towards}
Yupeng Hou, Shanlei Mu, Wayne~Xin Zhao, Yaliang Li, Bolin Ding, and Ji-Rong Wen. 2022.
\newblock Towards universal sequence representation learning for recommender systems.
\newblock In \emph{SIGIR}, pages 585--593.

\bibitem[{Li et~al.(2022)Li, Shang, and McAuley}]{li2022uctopic}
Jiacheng Li, Jingbo Shang, and Julian McAuley. 2022.
\newblock Uctopic: Unsupervised contrastive learning for phrase representations and topic mining.
\newblock \emph{arXiv preprint arXiv:2202.13469}.

\bibitem[{Li et~al.(2021)Li, Zhang, and Chen}]{li2021personalized}
Lei Li, Yongfeng Zhang, and Li~Chen. 2021.
\newblock Personalized transformer for explainable recommendation.

\bibitem[{Li et~al.(2023)Li, Zhang, and Chen}]{li2023personalized}
Lei Li, Yongfeng Zhang, and Li~Chen. 2023.
\newblock Personalized prompt learning for explainable recommendation.
\newblock \emph{Transactions on Information Systems (TOIS)}, 41(4):1--26.

\bibitem[{Li et~al.(2017)Li, Wang, Ren, Bing, and Lam}]{li2017neural}
Piji Li, Zihao Wang, Zhaochun Ren, Lidong Bing, and Wai Lam. 2017.
\newblock Neural rating regression with abstractive tips generation for recommendation.
\newblock In \emph{SIGIR}, pages 345--354.

\bibitem[{Ni et~al.(2019)Ni, Li, and McAuley}]{ni2019justifying}
Jianmo Ni, Jiacheng Li, and Julian McAuley. 2019.
\newblock Justifying recommendations using distantly-labeled reviews and fine-grained aspects.
\newblock In \emph{EMNLP}, pages 188--197.

\bibitem[{Qin et~al.(2023)Qin, Wang, Zhang, and Zhu}]{qin2023disentangled}
Yijian Qin, Xin Wang, Ziwei Zhang, and Wenwu Zhu. 2023.
\newblock Disentangled representation learning with large language models for text-attributed graphs.
\newblock \emph{arXiv preprint arXiv:2310.18152}.

\bibitem[{Ren et~al.(2024)Ren, Wei, Xia, Su, Cheng, Wang, Yin, and Huang}]{ren2023representation}
Xubin Ren, Wei Wei, Lianghao Xia, Lixin Su, Suqi Cheng, Junfeng Wang, Dawei Yin, and Chao Huang. 2024.
\newblock Representation learning with large language models for recommendation.
\newblock In \emph{WWW}, pages 3464--3475.

\bibitem[{Ren et~al.(2023)Ren, Xia, Zhao, Yin, and Huang}]{ren2023disentangled}
Xubin Ren, Lianghao Xia, Jiashu Zhao, Dawei Yin, and Chao Huang. 2023.
\newblock Disentangled contrastive collaborative filtering.
\newblock In \emph{SIGIR}, pages 1137--1146.

\bibitem[{Sellam et~al.(2020)Sellam, Das, and Parikh}]{sellam2020bleurt}
Thibault Sellam, Dipanjan Das, and Ankur~P Parikh. 2020.
\newblock Bleurt: Learning robust metrics for text generation.
\newblock In \emph{ACL}.

\bibitem[{Wang et~al.(2023)Wang, Liang, Meng, Sun, Shi, Li, Xu, Qu, and Zhou}]{wang2023chatgpt}
Jiaan Wang, Yunlong Liang, Fandong Meng, Zengkui Sun, Haoxiang Shi, Zhixu Li, Jinan Xu, Jianfeng Qu, and Jie Zhou. 2023.
\newblock Is chatgpt a good nlg evaluator? a preliminary study.
\newblock In \emph{EMNLP}.

\bibitem[{Wang et~al.(2019)Wang, He, Wang, Feng, and Chua}]{wang2019neural}
Xiang Wang, Xiangnan He, Meng Wang, Fuli Feng, and Tat-Seng Chua. 2019.
\newblock Neural graph collaborative filtering.
\newblock In \emph{SIGIR}, pages 165--174.

\bibitem[{Xi et~al.(2023)Xi, Liu, Lin, Zhu, Chen, Tang, Zhang, Zhang, and Yu}]{xi2023towards}
Yunjia Xi, Weiwen Liu, Jianghao Lin, Jieming Zhu, Bo~Chen, Ruiming Tang, Weinan Zhang, Rui Zhang, and Yong Yu. 2023.
\newblock Towards open-world recommendation with knowledge augmentation from large language models.
\newblock \emph{arXiv preprint arXiv:2306.10933}.

\bibitem[{Xia et~al.(2023)Xia, Huang, Huang, Lin, Yu, and Kao}]{xia2023automated}
Lianghao Xia, Chao Huang, Chunzhen Huang, Kangyi Lin, Tao Yu, and Ben Kao. 2023.
\newblock Automated self-supervised learning for recommendation.
\newblock In \emph{WWW}, pages 992--1002.

\bibitem[{Yan et~al.(2023)Yan, He, Li, Zhang, and McAuley}]{yan2023personalized}
An~Yan, Zhankui He, Jiacheng Li, Tianyang Zhang, and Julian McAuley. 2023.
\newblock Personalized showcases: Generating multi-modal explanations for recommendations.
\newblock In \emph{SIGIR}, pages 2251--2255.

\bibitem[{Yang et~al.(2023{\natexlab{a}})Yang, Wu, Wu, Zhang, Hong, Zhang, Zhou, and Wang}]{yang2023generative}
Yonghui Yang, Zhengwei Wu, Le~Wu, Kun Zhang, Richang Hong, Zhiqiang Zhang, Jun Zhou, and Meng Wang. 2023{\natexlab{a}}.
\newblock Generative-contrastive graph learning for recommendation.
\newblock In \emph{SIGIR}, pages 1117--1126.

\bibitem[{Yang et~al.(2023{\natexlab{b}})Yang, Huang, Xia, and Huang}]{yang2023knowledge}
Yuhao Yang, Chao Huang, Lianghao Xia, and Chunzhen Huang. 2023{\natexlab{b}}.
\newblock Knowledge graph self-supervised rationalization for recommendation.
\newblock In \emph{KDD}, pages 3046--3056.

\bibitem[{Yao et~al.(2021)Yao, Yi, Cheng, Yu, Chen, Menon, Hong, Chi, Tjoa, Kang et~al.}]{yao2021self}
Tiansheng Yao, Xinyang Yi, Derek~Zhiyuan Cheng, Felix Yu, Ting Chen, Aditya Menon, Lichan Hong, Ed~H Chi, Steve Tjoa, Jieqi Kang, et~al. 2021.
\newblock Self-supervised learning for large-scale item recommendations.
\newblock In \emph{CIKM}, pages 4321--4330.

\bibitem[{Ying et~al.(2018)Ying, He, Chen, Eksombatchai, Hamilton, and Leskovec}]{ying2018graph}
Rex Ying, Ruining He, Kaifeng Chen, Pong Eksombatchai, William~L Hamilton, and Jure Leskovec. 2018.
\newblock Graph convolutional neural networks for web-scale recommender systems.
\newblock In \emph{KDD}, pages 974--983.

\bibitem[{Yuan et~al.(2021)Yuan, Neubig, and Liu}]{yuan2021bartscore}
Weizhe Yuan, Graham Neubig, and Pengfei Liu. 2021.
\newblock Bartscore: Evaluating generated text as text generation.
\newblock In \emph{NeurIPS}, volume~34, pages 27263--27277.

\bibitem[{Zhang et~al.(2019{\natexlab{a}})Zhang, Shi, Zhao, and King}]{zhang2019star}
Jiani Zhang, Xingjian Shi, Shenglin Zhao, and Irwin King. 2019{\natexlab{a}}.
\newblock Star-gcn: Stacked and reconstructed graph convolutional networks for recommender systems.
\newblock In \emph{IJCAI}.

\bibitem[{Zhang et~al.(2019{\natexlab{b}})Zhang, Yao, Sun, and Tay}]{zhang2019deep}
Shuai Zhang, Lina Yao, Aixin Sun, and Yi~Tay. 2019{\natexlab{b}}.
\newblock Deep learning based recommender system: A survey and new perspectives.
\newblock \emph{ACM Computing Surveys (CSUR)}, 52(1):1--38.

\bibitem[{Zhang et~al.(2020)Zhang, Kishore, Wu, Weinberger, and Artzi}]{zhang2019bertscore}
Tianyi Zhang, Varsha Kishore, Felix Wu, Kilian~Q Weinberger, and Yoav Artzi. 2020.
\newblock Bertscore: Evaluating text generation with bert.
\newblock In \emph{ICLR}.

\end{thebibliography}
